\documentclass[aps,prl,twocolumn,superscriptaddress,showpacs,floatfix]{revtex4}

\usepackage{graphicx}

\begin{document}

\title{Magnetic Ordering in Blocking Layer and 
Highly Anisotropic Electronic Structure of High-$T_{\rm c}$
Iron-based Superconductor Sr$_2$VFeAsO$_3$: LDA+$U$ Studies}

\author{Hiroki Nakamura}
\email[]{nakamura.hiroki@jaea.go.jp}
\author{Masahiko Machida}
\email[]{machida.masahiko@jaea.go.jp}
\affiliation{CCSE, Japan Atomic Energy Agency, , 6--9--3 Higashi-Ueno,
Taito-ku Tokyo 110--0015, Japan}
\affiliation{CREST (JST), 4--1--8 Honcho, Kawaguchi, Saitama 332--0012,
Japan}
\affiliation{JST, Transformative Research-Project on Iron Pnictides (TRIP), Chiyoda, Tokyo 102-0075, Japan}

\date{\today}

\begin{abstract}
We calculate electronic structures of a high-$T_{\rm c}$ iron-based superconductor Sr$_2$VFeAsO$_3$  
by LDA+$U$ method.
We assume a checker-board antiferromagnetic order on blocking layers including vanadium and 
strong correlation in $d$-orbits of vanadium through the Hubbard $U$. While the standard LDA 
brings about metallic blocking layers and complicated 
Fermi surface as in the previous literatures, our 
calculation changes the blocking layer into insulating one and 
the Fermi surface becomes quite similar to those of other iron-based superconductors.
Moreover, the appearance of the insulating blocking layers 
predicts high anisotropy on quasi-particle transports and 
new types of intrinsic Josephson effects.
\end{abstract}

\maketitle

Since the discovery of high-$T_{\rm c}$ superconductivity 
in LaFeAsO$_{1-x}$F$_x$ \cite{kamihara}, its various family members 
have been piled up on the materials table of iron-based superconductors.
The materials variety is characterized by 
their crystal structures and often classified 
by the numbering scheme, such as ``1111" ({\it e.g.},  LaFeAsO$_{1-x}$F$_x$),
``122" ({\it e.g.}, Ba$_{1-x}$K$_x$Fe$_2$As$_2$ \cite{122}),
``111" ({\it e.g.}, LiFeAs \cite{111}),
and ``11'' ({\it e.g.}, FeSe \cite{11}).
They have quasi-two-dimensional FeAs layers commonly,
while a main difference among them comes from the non-superconducting blocking layers between the FeAs layers.
The density functional theory with local density approximation (LDA) 
has predicted a common electronic structure, i.e., multiple cylindrical Fermi surfaces consisting of 
hole pockets around the Brillouin zone center and electron ones around the zone corners.
Some experiments, e.g., ARPES's have actually confirmed such an electronic structure \cite{arpes}.
In this case, a strong candidate of the glue of their superconducting pairs has been regarded to be  
a spin fluctuation 
due to the nesting between those Fermi pockets
\cite{kuroki}, though it still remains unsettled.

Recently, a new family who has a thick
perovskite-type blocking layer has been discovered \cite{ogino} and 
a controversial debate about the pairing mechanism has arisen \cite{lee,mazin,shein}.
These materials include 
non-iron transition-metal elements in their thick blocking layers, and 
the electronic structures can be relatively more complex than the other types, if 3$d$ orbitals 
on the transition metals hybridize with Fe $d$-orbitals.
Such hybridization can clearly break the typical stage
composed of disconnected small hole and electron pockets.
Sr$_2$VFeAsO$_3$ is one of such perovskite-type iron-based superconductors,
who exhibits the highest $T_{\rm c}\sim 37$K in this family \cite{zhu}.
Initially, standard LDA calculations revealed much different and more complex structures 
due to the hybridization of the 3$d$ orbitals of vanadium.
Thus, the LDA result 
gave rise to the following two interpretations.
Some reports suggested that the pairing is not originated from the nested Fermi surfaces \cite{lee}, while 
the nesting feature is still alive even in such a complex structure \cite{mazin}.
Thus, Sr$_2$VFeAsO$_3$ can be regarded as a key material to check 
whether the typical Fermi-surface structure is essential for high-$T_{\rm c}$ superconductivity.
Therefore, we examine whether
the electronic structure of Sr$_2$VFeAsO$_3$ is really different from the typical one
in ``conventional" iron-based superconductors.

Vanadium oxide is well known to be a strongly correlated system.
For instance, V$_2$O$_3$ exhibits a phase transition into a Mott insulator at a certain temperature,
though standard LDA calculations predict its metallic features \cite{ezhov}.
On the other hand, the calculations considering the strong correlation such as LDA+$U$ \cite{ezhov} and LDA+DMFT \cite{held}
succeeded in reproducing the Mott insulating state.
The nominal valence of V in V$_2$O$_3$ is trivalent, and  
V in Sr$_2$VFeAsO$_3$ is naively estimated to be also trivalent from the charge 
valence.
This implies that the perovskite-layer including vanadium oxide becomes insulating
and then the Fermi surfaces are not influenced by vanadium electrons.
In this letter, we explore the electronic structures of Sr$_2$VFeAsO$_3$ 
by considering the correlation on 3$d$ vanadium electrons.
For this purpose, we use LDA+$U$ method for $d$-electrons of vanadium.
As a result, we find that the calculated electronic structure 
becomes equivalent to those of other iron-based superconductors
as expected above, by assuming
a checker-board antiferromagnetic (AFM) order on vanadium layer.
In this case, highly anisotropic electronic structures are also obtained
in contrast to the standard calculation results.

\begin{figure}
\includegraphics[width=8.3cm]{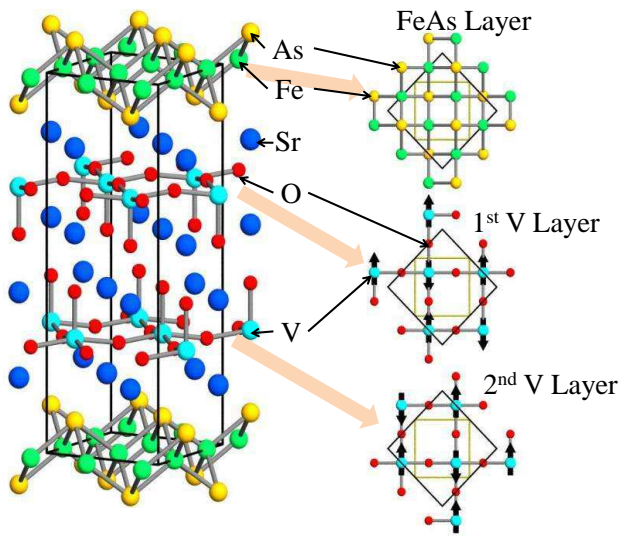}
\caption{Crystal structure of Sr$_2$VFeAsO$_3$.
The right panel displays each two-dimensional layer seen from $c$-axis.
The black small arrows on each V layer denote spin directions in the assumed checker-board antiferromagnetic order.
The small squares in the right panel stand for original unit cell (space group: $P4/nmm$).
Thick lines denotes the extended unit cell for checker-AFM.
}\label{fig1}
\end{figure}

The crystal structure of the present target material, 
Sr$_2$VFeAsO$_3$ is tetragonal, and the space group of 
the crystal structure is $P4/nmm$ as shown in Fig.~\ref{fig1}. 
Such a structure is characteristic to the perovskite-type iron-based superconductors
but in contrast to most of other family materials whose mother 
compounds without doping 
undertake the structural transition 
into the orthogonal one.
This non-doped high-$T_{\rm c}$ superconductor without
the orthogonal transition is relatively convenient for first-principles studies
because we do not need to care any doping effect.

We consider strong correlation on vanadium $d$-orbital electrons.
This is the first trial for 
the present compound. 
We point out that it is only a way to reproduce some experimental results consistently. 
In order to include the strong correlation effect, we use LDA + $U$ method, in which 
the Hubbard $U$ is applied only on the vanadium $d$-electrons.
In our LDA + $U$ calculation, we expect a checker-AFM order on the vanadium layer 
as a natural consequence of significantly large $U$. 
Then, the unit cell is extended to $\sqrt{2}a\times\sqrt{2}a$ in $ab$ plane (see Fig.~\ref{fig1}).
As a result, the Brillouin zone is folded and
the zone corner called $M$-point coincides with the zone center ($\Gamma$-point).
In the calculation unit cell, there are two vanadium-layers where
 checker-AFM orders are arranged as described in Fig.1. 
The magnetic order is equivalent to the observed one 
in Sr$_2$CrFeAsO$_3$ \cite{Cr}, which does not exhibit superconductivity at all.
The difference between these materials is an interesting issue, which will 
be discussed elsewhere. 
On the other hand, we do not set any magnetic order 
for the iron layers except for a case to compare the total energy.
The calculation package employed throughout this paper is VASP \cite{vasp, calc}
that supports LDA+U method \cite{ldapu},
in which we choose two parameters, the Hubbard $U$ and Hund's coupling $J$.
While there are methods to calculate them in first-principles manner, e.g., the constrained RPA \cite{imada},
we treat them as input parameters conventionally.
Instead, we examine how the electronic structure is affected by various parameter sets.
We find that a combination of $U=5.5$ eV and $J=0.93$ eV is 
the most realistic among all sets used in this paper.
The set is close to those employed in 
 successful calculations on vanadium oxides using LDA + $U$ and LDA+DMFT \cite{held,solovyev}. 
The lattice constants and internal coordinate of each atom are in accordance with 
the experimental data \cite{zhu}.

\begin{figure}
\begin{flushleft}
(a) \\
\includegraphics[width=8.3cm]{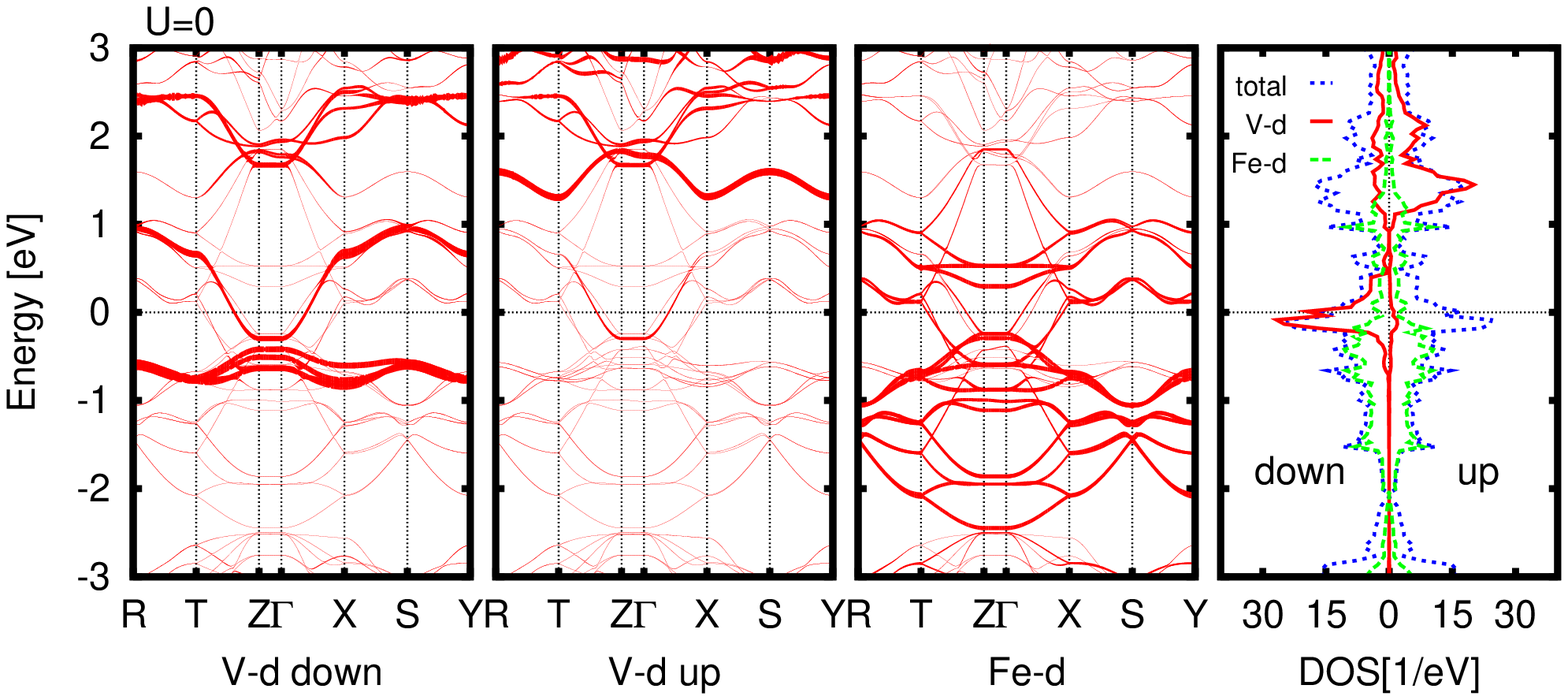} \\
(b) \\
\includegraphics[width=8.3cm]{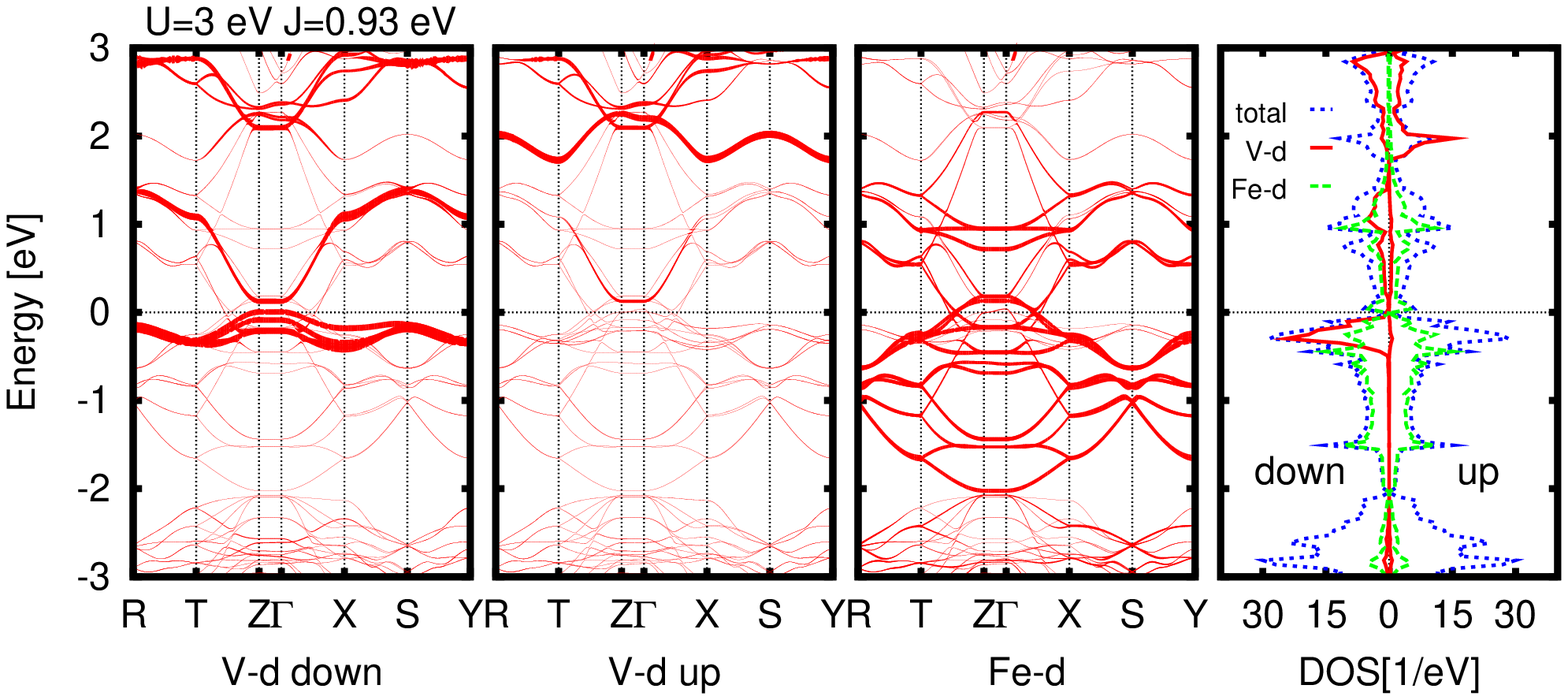} \\
(c) \\
\includegraphics[width=8.3cm]{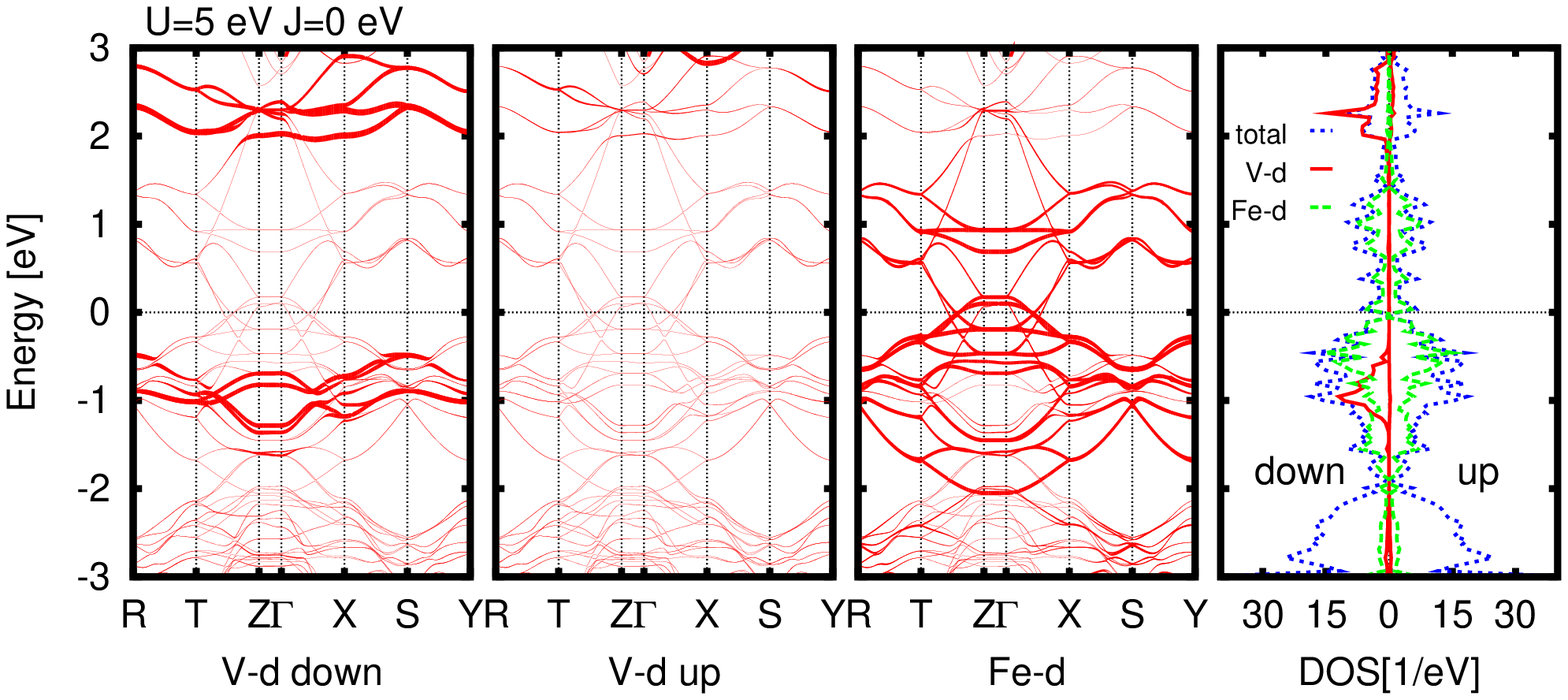} \\
(d) \\
\includegraphics[width=8.3cm]{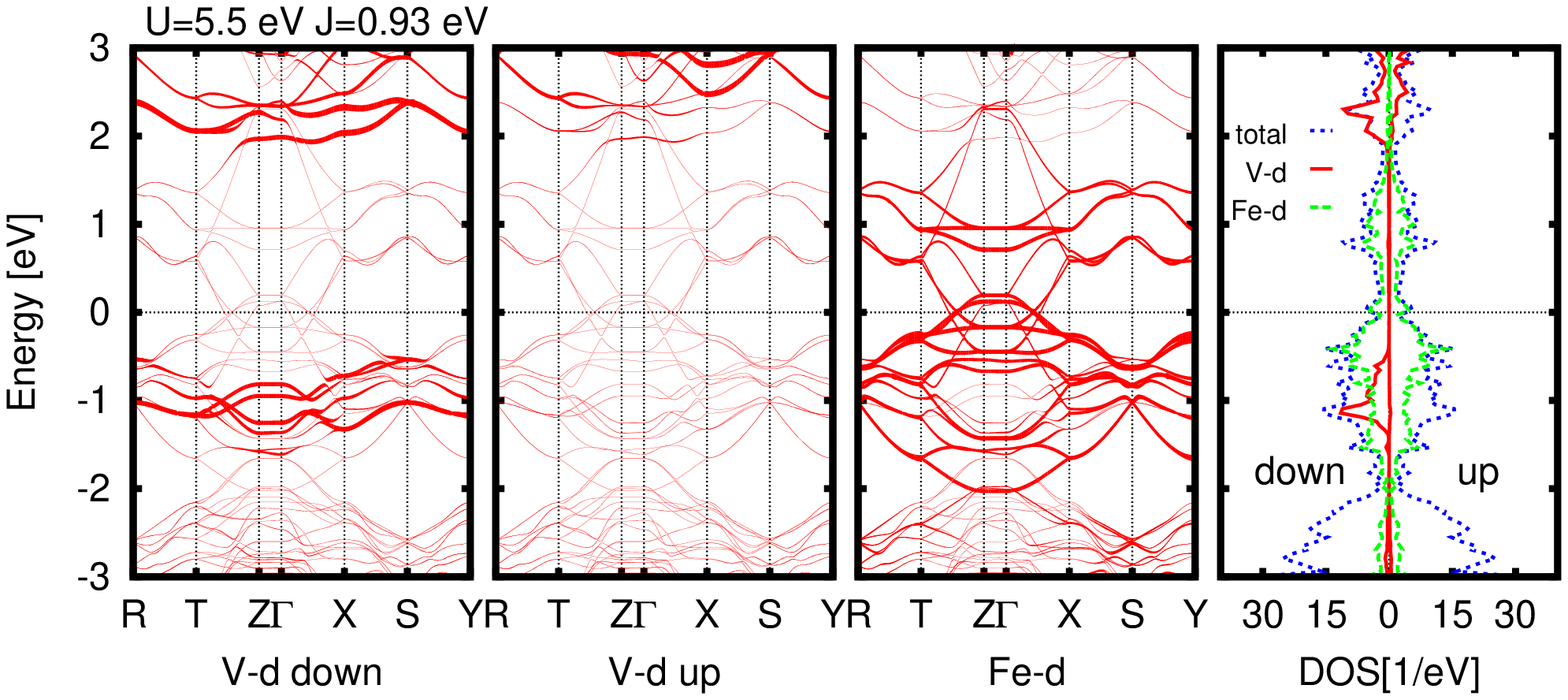} \\
\end{flushleft}
\caption{Band structure and density of states calculated by LDA+U for
(a) $U=0$, $J=0$, (b) $U=3$ eV, $J=0.93$ eV, (c) $U=5$ eV, $J=0$ eV, and
(d) $U=5.5$ eV, $J=0.93$ eV. 
The contributions of V-$d$ orbital spin-down, spin-up, and Fe-$d$ orbitals to the
band dispersions are, respectively, highlighted by bold-lines from the left-hand-side panel 
to the right-hand, in (a) to (d).
The total and partial Fe, and partial V DOS's are depicted by the dotted, dashed, and solid curves, respectively,
in the right-end panel. 
}
\label{fig2}
\end{figure}

Figure \ref{fig2} shows band dispersions and densities of states calculated by
LDA+U method for the four parameter sets, (a)$U=0$,$J=0$, (b)$U=3$ eV, $J=0.93$ eV,
(c) $U=5$ eV, $J=0$ (d) $U=5.5$ eV, $J=0.93$ eV.
In all sets of $U$ and $J$, the checker-AFM order in vanadium layer is stable,
and the calculated moments on vanadium become 1.36, 1.65, 1.86, and 1.79 $\mu_B$ for the parameter sets (a) to (d),
respectively.
In $U=0$, that is, a standard LDA calculation, 
the bands of spin-down (majority-spin) $d$-electron of vanadium
cross the Fermi level as shown in Fig.~\ref{fig2}(a),
though spin-up bands are away from the Fermi level.
In this case, the perovskite layer is not insulating but half-metallic as pointed out in
 Ref.~\cite{shein}.
As $U$ increases, the occupied and unoccupied bands of vanadium split and both of those go away from the Fermi level.
For reasonable values, i.e., $U=5$ eV and $U=5.5$ eV, they clearly split into upper and 
lower Hubbard bands, and the blocking layer becomes insulating.

\begin{figure}
\includegraphics[width=8.3cm]{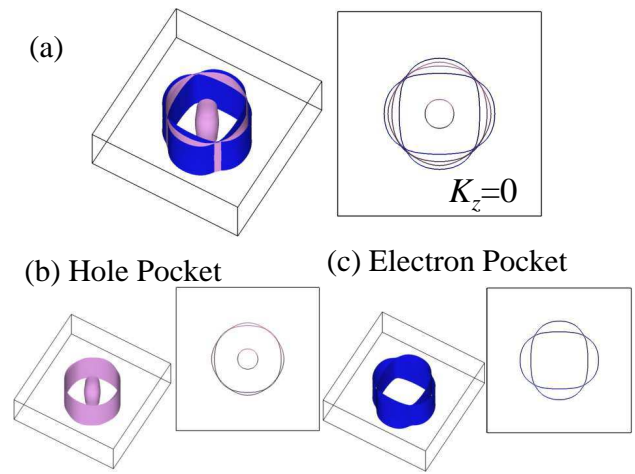}
\caption{Fermi Surfaces for $U=5.5$ eV and $J=$0.93 eV.
(a) all Fermi surfaces. The dark and light surfaces stand for
electron and hole pockets, respectively.
The right panel shows a sliced Fermi surface at $k_z=0$.
(b) and (c)  describe Fermi surfaces separately for hole and electron pockets, respectively.
}\label{fig3}
\end{figure}

The Fermi surfaces for the set of $U=5.5$ eV and $J=0.93$ eV are shown in Fig.~\ref{fig3}.
The surface is formed by only $d$-electrons of iron, since the perovskite layer including vanadium 
is not metallic any longer.
In the figure, one finds that all Fermi surfaces are cylindrical around $\Gamma$-point
and two of them are formed as hole pockets and other three as electron ones.
We note that the present Brillouin zone is folded because of the checker-AFM ordering.
As a consequence of the folding, the electron pockets seen in the zone center
correspond to those around $M$-point calculated in
other iron-based superconducting materials.
Thus, we find that the calculated Fermi surfaces are quite equivalent to 
those of other typical iron-based superconductors.
Moreover, our folded result clearly shows that the nesting between those hole and electron pockets is well and 
the spin fluctuation with the nesting vector is expected to be strongly enhanced.
Very recently, Nakayama et al. actually reported typical Fermi surfaces as predicted above by using ARPES \cite{tohoku}.
This is strong evidence that vanadium 3$d$ electrons do not contribute to  
band structure around the Fermi level, i.e., the vanadium layer is insulating.  

In the case of $U=5.5$ eV and $J=0.93$ eV, we also calculate anisotropy of quasi-particle resistivity.
Some experimental data qualitatively indicates that the anisotropy is quite large \cite{zhu2}. 
One of the clearest evidence is the broadening of the superconducting transition 
by the variation of the applied magnetic field, which 
is well known on highly anisotropic high-$T_{\rm c}$ cuprate materials such as Bi$_2$Sr$_2$CaCu$_2$O$_x$ (Bi-2212).
The calculated anisotropy of the Fermi velocity, $\gamma_\rho ( = \langle v_a^2 \rangle /\langle v_c^2 \rangle)$
becomes 1351 which is much larger than those of LaFeAsO ($\gamma_\rho=116.8$) 
and BaFe$_2$As$_2$ ($\gamma_\rho=10.69$), but relatively smaller than
perovskite-type Sr$_2$ScFePO$_3$ ($\gamma_\rho=6.19\times 10^5$) \cite{nakamura}. 
The anisotropy of the penetration depth $\gamma_\lambda$ at zero temperature is $\sim$ 37, since
 $\gamma_\lambda=\sqrt{\gamma_\rho}$ at zero temperature.
Although the calculated anisotropy is relatively smaller among the family of perovskite-type
iron-based superconductors, it is still much larger than other families.
Compared to those of high-$T_{\rm c}$ cuprate superconductors, the anisotropy $\gamma_\lambda \sim 30$ is
larger than YBa$_2$Cu$_3$O$_{7-x}$ and comparable to La$_{2-x}$Sr$_x$CuO$_4$ but smaller than Bi-2212.
The value is fully over the lower bound anisotropy, in which intrinsic Josephson effects are observable.
In fact, since intrinsic Josephson effects have been confirmed in 1111 systems \cite{1111josephson} whose 
$\gamma_\lambda \sim 10$ according to our first-principles calculations, the expectation 
is reasonable. 
Thus, intrinsic Josephson effects are promised.
Moreover, we can predict some new effects originated from its multi-band superconducting gaps in the 
intrinsic Josephson junction systems.
Some literatures report a new excitation called Josephson-Leggett mode in addition to
the Josephson plasma due to multi-tunneling channels\cite{josephson1}, 
and the others predict new type of Josephson effects originated 
from the multi-degree of freedom of the superconducting phases \cite{josephson2}.
At the present, Bi-2212 is intensively investigated to promote the functionality as THz wave radiation source.
The single crystal of Sr$_2$VFeAsO$_3$ may work as multi-THz wave generators due to its multiple excitation modes.   

Finally, let us discuss stability of the present checker-AFM state and mention the related issues.
First, we compare the total energy of the checker-AFM magnetic state 
with that of non-magnetic one. The both are calculated by LDA+U with the parameter set, $U=5.5$ eV and $J=0.93$ eV.
In our calculation, the total energy per formula of the non-magnetic state is 
2 eV larger than that of the checker-AFM state.
This result clearly indicates that the checker-AFM state is more stable than non-magnetic one under the 
application of the Hubbard $U$ on the vanadium $d$-electrons.
Very recently, Tatematsu et al., reported that a magnetic transition occurs around 150K at the vanadium
layers by NMR measurement \cite{nagoya}. 
Although the transition is not yet proved to be the checker-AFM ordering, it  
can be ascribed to a magnetic ordering at vanadium layer at least.
Otherwise, the transition is attributed to Fe stripe ordering but such ordering 
drastically changes the Fermi surfaces, which is not consistent with ARPES measurement \cite{tohoku}.
On the other hand, we have another interesting calculation result, in which
the total energy of Fe-stripe AFM together with the checker-AFM of vanadium is 0.2 eV smaller than that 
of non-magnetic Fe with the checker-AFM of vanadium.
Then, the calculated Fe magnetic moment is 2.1 $\mu_B$, which is comparable to those 
calculated in most of mother compounds of iron-based superconductors.
It is well known that LDA calculations always predict that the stripe-AFM state with Fe moment $\sim 2\mu_{\rm B}$ is stable 
even in the superconducting doping range.
This result indicates that the stripe-AFM instability on the iron-layer strongly works 
even in Sr$_2$VFeAsO$_3$ similar to other compounds. 
Such an agreement is not an accident but a clear evidence for the magnetic instability on 
Fe layers common to all families of iron-based superconductors.

In conclusions, we calculate the electronic structure of Sr$_2$VFeAsO$_3$ using LDA+$U$ scheme under an assumption of 
the checker-AFM ordering on vanadium-layers.
A reasonable choice of the parameter $U$ and $J$ leads to 
the insulating blocking layer
and typical cylindrical Fermi surface structures.
In addition, our results predict high anisotropy as observed in recent experiments. 
Consequently, we conclude that the electronic structures are quite equivalent to those of other 
iron-based superconductors and the similar magnetic instability may contribute to the pairing. 
Furthermore, the high-anisotropy in superconducting transport properties 
predicts new types of intrinsic Josephson 
effects originating from the multi-band superconductivity.
The present result can settle down the controversial problem for the pairing mechanism and 
suggests a new application possibility.

The authors wish to thank  K. Terakura and
 N. Hamada for illuminating discussions in first-principles calculations,
 and T. Sato, Y. Kobayashi and Y. Matsuda for stimulating discussions about experimental results.
 The authors also thank H. Aoki,
 N. Hayashi, Y. Nagai, M. Okumura, R. Igarashi, Y. Ota and N. Nakai 
 for valuable discussion. The work was partially supported by Grant-in-Aid 
 for Scientific Research on Priority Area "Physics of new quantum phases 
 in superclean materials" (Grant No. 20029019) from the Ministry of 
 Education, Culture, Sports, Science and Technology of Japan.


\begin{thebibliography}{99}
\bibitem{calc}
In our calculations, we use the PAW method \cite{paw}
with GGA exchange-correlation energy \cite{pbe}.
K-points are taken as $8\times 8\times 3$ for self-consistent loop
and $10 \times 10 \times 4 $ for density of states.
Self-consistent loop is repeated until the energy difference between loops
becomes less than $10^{-7}$ eV.
\end{thebibliography}
\end{document}